\documentclass[final]{IEEEtran}
\usepackage{amsthm,amssymb,graphicx,multirow,amsmath,color,amsfonts,physics}
\usepackage[update,prepend]{epstopdf}
\usepackage[noadjust]{cite}
\usepackage{tikz}
\usepackage{bbm} 
\usepackage{pdfpages}
\usepackage{balance}
\usepackage{multirow}
\usepackage{comment}
\usepackage{subfigure}

\allowdisplaybreaks 

\begin{document}
\def\nba{{\mathbf{a}}}
\def\nbb{{\mathbf{b}}}
\def\nbc{{\mathbf{c}}}
\def\nbd{{\mathbf{d}}}
\def\nbe{{\mathbf{e}}}
\def\nbf{{\mathbf{f}}}
\def\nbg{{\mathbf{g}}}
\def\nbh{{\mathbf{h}}}
\def\nbi{{\mathbf{i}}}
\def\nbj{{\mathbf{j}}}
\def\nbk{{\mathbf{k}}}
\def\nbl{{\mathbf{l}}}
\def\nbm{{\mathbf{m}}}
\def\nbn{{\mathbf{n}}}
\def\nbo{{\mathbf{o}}}
\def\nbp{{\mathbf{p}}}
\def\nbq{{\mathbf{q}}}
\def\nbr{{\mathbf{r}}}
\def\nbs{{\mathbf{s}}}
\def\nbt{{\mathbf{t}}}
\def\nbu{{\mathbf{u}}}
\def\nbv{{\mathbf{v}}}
\def\nbw{{\mathbf{w}}}
\def\nbx{{\mathbf{x}}}
\def\nby{{\mathbf{y}}}
\def\nbz{{\mathbf{z}}}
\def\nb0{{\mathbf{0}}}
\def\nb1{{\mathbf{1}}}

\def\nbA{{\mathbf{A}}}
\def\nbB{{\mathbf{B}}}
\def\nbC{{\mathbf{C}}}
\def\nbD{{\mathbf{D}}}
\def\nbE{{\mathbf{E}}}
\def\nbF{{\mathbf{F}}}
\def\nbG{{\mathbf{G}}}
\def\nbH{{\mathbf{H}}}
\def\nbI{{\mathbf{I}}}
\def\nbJ{{\mathbf{J}}}
\def\nbK{{\mathbf{K}}}
\def\nbL{{\mathbf{L}}}
\def\nbM{{\mathbf{M}}}
\def\nbN{{\mathbf{N}}}
\def\nbO{{\mathbf{O}}}
\def\nbP{{\mathbf{P}}}
\def\nbQ{{\mathbf{Q}}}
\def\nbR{{\mathbf{R}}}
\def\nbS{{\mathbf{S}}}
\def\nbT{{\mathbf{T}}}
\def\nbU{{\mathbf{U}}}
\def\nbV{{\mathbf{V}}}
\def\nbW{{\mathbf{W}}}
\def\nbX{{\mathbf{X}}}
\def\nbY{{\mathbf{Y}}}
\def\nbZ{{\mathbf{Z}}}

\def\ncalA{{\mathcal{A}}}
\def\ncalB{{\mathcal{B}}}
\def\ncalC{{\mathcal{C}}}
\def\ncalD{{\mathcal{D}}}
\def\ncalE{{\mathcal{E}}}
\def\ncalF{{\mathcal{F}}}
\def\ncalG{{\mathcal{G}}}
\def\ncalH{{\mathcal{H}}}
\def\ncalI{{\mathcal{I}}}
\def\ncalJ{{\mathcal{J}}}
\def\ncalK{{\mathcal{K}}}
\def\ncalL{{\mathcal{L}}}
\def\ncalM{{\mathcal{M}}}
\def\ncalN{{\mathcal{N}}}
\def\ncalO{{\mathcal{O}}}
\def\ncalP{{\mathcal{P}}}
\def\ncalQ{{\mathcal{Q}}}
\def\ncalR{{\mathcal{R}}}
\def\ncalS{{\mathcal{S}}}
\def\ncalT{{\mathcal{T}}}
\def\ncalU{{\mathcal{U}}}
\def\ncalV{{\mathcal{V}}}
\def\ncalW{{\mathcal{W}}}
\def\ncalX{{\mathcal{X}}}
\def\ncalY{{\mathcal{Y}}}
\def\ncalZ{{\mathcal{Z}}}

\def\nbbA{{\mathbb{A}}}
\def\nbbB{{\mathbb{B}}}
\def\nbbC{{\mathbb{C}}}
\def\nbbD{{\mathbb{D}}}
\def\nbbE{{\mathbb{E}}}
\def\nbbF{{\mathbb{F}}}
\def\nbbG{{\mathbb{G}}}
\def\nbbH{{\mathbb{H}}}
\def\nbbI{{\mathbb{I}}}
\def\nbbJ{{\mathbb{J}}}
\def\nbbK{{\mathbb{K}}}
\def\nbbL{{\mathbb{L}}}
\def\nbbM{{\mathbb{M}}}
\def\nbbN{{\mathbb{N}}}
\def\nbbO{{\mathbb{O}}}
\def\nbbP{{\mathbb{P}}}
\def\nbbQ{{\mathbb{Q}}}
\def\nbbR{{\mathbb{R}}}
\def\nbbS{{\mathbb{S}}}
\def\nbbT{{\mathbb{T}}}
\def\nbbU{{\mathbb{U}}}
\def\nbbV{{\mathbb{V}}}
\def\nbbW{{\mathbb{W}}}
\def\nbbX{{\mathbb{X}}}
\def\nbbY{{\mathbb{Y}}}
\def\nbbZ{{\mathbb{Z}}}

\def\nfrakR{{\mathfrak{R}}}

\def\nrma{{\rm a}}
\def\nrmb{{\rm b}}
\def\nrmc{{\rm c}}
\def\nrmd{{\rm d}}
\def\nrme{{\rm e}}
\def\nrmf{{\rm f}}
\def\nrmg{{\rm g}}
\def\nrmh{{\rm h}}
\def\nrmi{{\rm i}}
\def\nrmj{{\rm j}}
\def\nrmk{{\rm k}}
\def\nrml{{\rm l}}
\def\nrmm{{\rm m}}
\def\nrmn{{\rm n}}
\def\nrmo{{\rm o}}
\def\nrmp{{\rm p}}
\def\nrmq{{\rm q}}
\def\nrmr{{\rm r}}
\def\nrms{{\rm s}}
\def\nrmt{{\rm t}}
\def\nrmu{{\rm u}}
\def\nrmv{{\rm v}}
\def\nrmw{{\rm w}}
\def\nrmx{{\rm x}}
\def\nrmy{{\rm y}}
\def\nrmz{{\rm z}}

\def\nbydef{:=}
\def\nborel{\ncalB(\nbbR)}
\def\nboreld{\ncalB(\nbbR^d)}
\def\sinc{{\rm sinc}}

\newtheorem{lemma}{Lemma}
\newtheorem{thm}{Theorem}
\newtheorem{definition}{Definition}
\newtheorem{ndef}{Definition}
\newtheorem{nrem}{Remark}
\newtheorem{theorem}{Theorem}
\newtheorem{prop}{Proposition}
\newtheorem{cor}{Corollary}
\newtheorem{example}{Example}
\newtheorem{remark}{Remark}
\newtheorem{assumption}{Assumption}
	

\newcommand{\ceil}[1]{\lceil #1\rceil}
\def\argmin{\operatorname{arg~min}}
\def\argmax{\operatorname{arg~max}}
\def\figref#1{Fig.\,\ref{#1}}%
\def\E{\mathbb{E}}
\def\EE{\mathbb{E}^{!o}}
\def\P{\mathbb{P}}
\def\pc{\mathtt{P_c}}
\def\rc{\mathtt{R_c}}   
\def\p{p}

\def\V{\operatorname{Var}}
\def\erfc{\operatorname{erfc}}
\def\erf{\operatorname{erf}}
\def\opt{\mathrm{opt}}
\def\R{\mathbb{R}}
\def\Z{\mathbb{Z}}

\def\LL{\mathcal{L}^{!o}}
\def\var{\operatorname{var}}
\def\supp{\operatorname{supp}}

\def\N{\sigma^2}
\def\T{\beta}							
\def\sinr{\mathtt{SINR}}			
\def\snr{\mathtt{SNR}}
\def\sir{\mathtt{SIR}}
\def\ase{\mathtt{ASE}}
\def\se{\mathtt{SE}}

\def\calN{\mathcal{N}}
\def\FE{\mathcal{F}}
\def\calA{\mathcal{A}}
\def\calK{\mathcal{K}}
\def\calT{\mathcal{T}}
\def\calB{\mathcal{B}}
\def\calE{\mathcal{E}}
\def\calP{\mathcal{P}}
\def\calL{\mathcal{L}}


\def\l{\ell}
\newcommand{\fad}[2]{\ensuremath{\mathtt{h}_{#1}[#2]}}
\newcommand{\h}[1]{\ensuremath{\mathtt{h}_{#1}}}

\newcommand{\err}[1]{\ensuremath{\operatorname{Err}(\eta,#1)}}
\newcommand{\FD}[1]{\ensuremath{|\mathcal{F}_{#1}|}}



\def\Bx{{\mathcal{B}}^x}
\def\Bxx{{\mathcal{B}}^{x_0}}
\def\jx{y}
\def\m{(\bar{n}-1)}
\def\mm{\bar{n}-1}
\def\Nx{{\mathcal{N}}^x}
\def\Nxo{{\mathcal{N}}^{x_0}}
\def\wj{w_{jx_0}}
\def\uij{u_{jx}}
 \def\yj{y}
 \def\yjx{y}
 \def\zjx{z_x}
 \def \tx {y_0}
 \def \htx {h_0}

\def\rx{z_{1}}
\def\ry{z_{2}}

\def\Rx{Z_{1}}
\def\Ry{Z_{2}}

\def \hyxx {h_{y_{x_0}}}
\def \hyx {h_{y_x}}

\def\nbb1{\mathbbm{1}}
\def\xi{x_i}
\def\xj{x_j}
\def\xx{x_0}
\def\yk{y_k}
\def\yy{y_0}
\def\ie{{\em i.e. }}
\def\eg{{\em e.g. }}
\def\iid{{\em i.i.d. }}
\def\avg{\rm avg}

\def\rmnuma{\rm\uppercase\expandafter{\romannumeral1}}
\def\rmnumb{\rm\uppercase\expandafter{\romannumeral2}}
\def\rmnumc{\rm\uppercase\expandafter{\romannumeral3}}
\def\rmnumd{\rm\uppercase\expandafter{\romannumeral4}}
\def\rmnume{\rm\uppercase\expandafter{\romannumeral5}}
\def\rmnumf{\rm\uppercase\expandafter{\romannumeral6}}
\pagenumbering{gobble}
\graphicspath{{./Figures/}}
\title{
Virtual Backhaul Connectivity for Enhanced Coverage in Fiber-less Areas}
\author{
 Hao Lin,~\IEEEmembership{Graduate Student Member,~IEEE}, Mustafa A. Kishk,~\IEEEmembership{Member,~IEEE} \\and Mohamed-Slim Alouini,~\IEEEmembership{Fellow,~IEEE}
\thanks{Hao Lin is with the Electrical and Computer Engineering Program, CEMSE Division, King Abdullah University of Science and Technology (KAUST),
Thuwal 23955-6900, Kingdom Saudi Arabia (e-mail: hao.lin.std@gmail.com).\\
\indent Mustafa A. Kishk is with the Department of Electronic Engineering,
Maynooth University, Maynooth, W23 F2H6 Ireland (e-mail:
mustafa.kishk@mu.ie).\\
\indent Mohamed-Slim Alouini is with the CEMSE Division, King Abdullah
University of Science and Technology (KAUST), Thuwal 23955-6900,
Saudi Arabia (e-mail: slim.alouini@kaust.edu.sa).}
}

\maketitle
\begin{abstract}
This paper provides an overview of potential alternatives for providing wireless backhaul in regions that suffer from the lack of fiber optic-connectivity to the core network. These regions can be (i) rural and remote locations, (ii) low-income neighborhoods in urban and suburban regions, and (iii) post-disaster locations suffering from the destruction of cellular infrastructure. For these scenarios, extending fiber optic cables to such locations might be extremely expensive, impractical, or simply not feasible. Hence, in order to enhance the backhaul connectivity in these scenarios, we study the potential and applicability of the integrated access and backhaul (IAB) technique, and a hybrid combination of IAB and non-terrestrial networks (NTN) that includes high/low altitude platforms (HAPs/LAPs) and low earth orbit (LEO) satellites. We conclude this article by discussing the design considerations and potential research problems that would enable efficient deployment of such solutions.\end{abstract}
\begin{IEEEkeywords}
Integrated access and backhaul, fiber-less regions, non-terrestrial networks, digital divide
\end{IEEEkeywords}

\section*{Introduction} \label{sec:intro}
\indent 
Internet connectivity has become a vital part of many services such as remote learning, smart health care, and modern agriculture. {However, in the whole world, more than half of the population still lacks proper internet connectivity, especially in rural and remote areas. This, in turn, limits the opportunities provided to inhabitants of these regions to enhance their quality of life. Such an imbalance of quality of Internet services between underserved areas and developed areas is called \textit{the digital divide} \cite{chaoub20216g}. In the 5G era, more Sustainable Development Goals, including quality education for remote areas, sustainable agriculture, good health and well-being, full and productive employment, and decent work in less developed regions, are proposed to bridge such a digital divide around the world.} \\
\indent Optical fiber, as a method to achieve high-speed and long-distance connection, has become one of the important technologies for building communication networks. For example, in poverty-stricken areas of China, by 2020, the proportion of optical fiber coverage has increased to 98\%. Fiber-to-the-home (FTTH) population coverage reached 55.6\% in Europe in 2022, up from 50\% in 2021. However, in most developing countries, fiber optic connections still face difficulties and challenges. FTTH Council Europe's reports show that by September 2021, only 30\% of rural inhabitants could enjoy capabilities offered by full-fiber connectivity \cite{ETNO2023}. There are many reasons for the lack of optical fiber: the cost of optical fiber, the complexity of local terrain, and the design of the fiber network. Hence, the coverage of communication networks in rural and remote areas is much weaker than that in developed regions, \ie \textit{the digital divide from the fiber optic perspective}.\\
\indent {In order to bridge the digital divide, the next-generation communication aims to provide reliable gigabit broadband services to more people in different regions. Internet service providers have two main choices: optical fiber or wireless technology. In recent years, the 3rd Generation Partnership Project (3GPP) has developed 5G New Radio (NR) to enable people to use fiber-like high-speed network services faster and more efficiently. In 5G NR, operators can divide the wide bandwidth into two parts for wireless backhaul and access services, and this technique is known as integrated access and backhaul (IAB) network \cite{polese2020integrated,zhang2021survey}. For example, WeLink and Meta's Terragraph adopt the IAB technology to provide backhaul services for fixed wireless access (FWA) using millimeter wave (mmWave) channels. Since it can provide wireless backhaul service at fiber-like speeds, IAB technology is also known as ``virtual fiber". However, such IAB networks still require fiber Point-of-Presence nodes (PoPs) to connect to the provider's backbone network. Due to the high expense of fiber cable, the spatial constraints of fiber PoPs make it difficult for the IAB network in rural or remote areas to realize its potential.} \\
\indent {Since the approval in Release 15 in March 2017, 3GPP has proposed many solutions to integrate non-terrestrial (NTN) networks into the 5G NR system, due to its potential to realize wider coverage and higher throughput \cite{NTN}. The NTN technologies, including low earth orbit (LEO) satellites and high/low altitude platforms (HAPs/LAPs), have been applied for emergency response, backhauling, data collection and IoT in remote areas. Therefore, it can be a feasible solution to utilize NTN platforms to overcome the challenges of applying IAB networks in areas lacking optical fiber.}
\section*{Examples of fiber-less regions} \label{sec:example}
\begin{figure*}
    \centering
    \includegraphics[width=1.2\columnwidth]{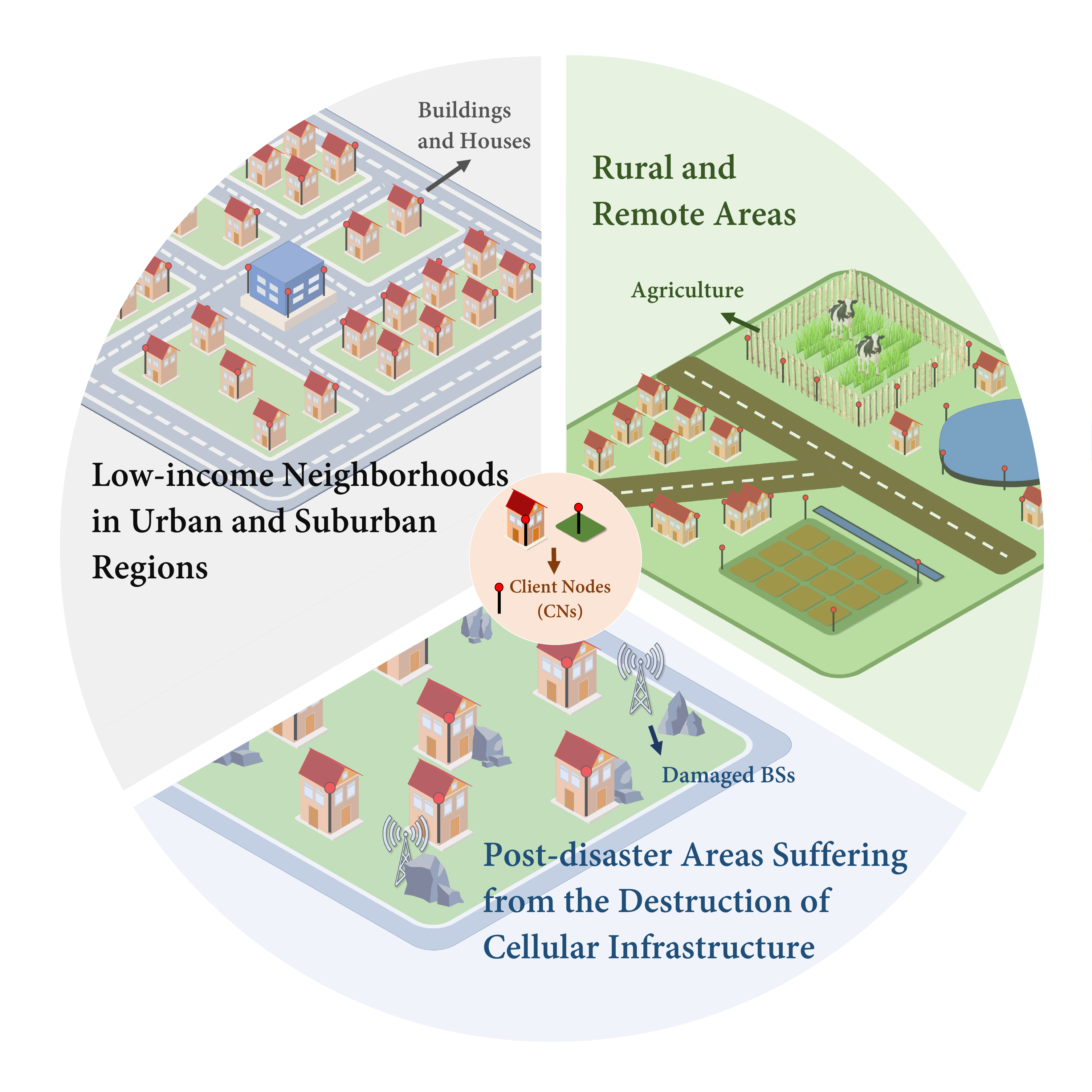}
    \caption{Three use cases of Integrated Access and Backhaul networks: Smart agriculture, urban construction, and post-disaster reconstruction are all sources of demand for backhaul services and CPEs (such as Wi-Fi APs). The client nodes (CNs) are installed in front of buildings, houses, farmland, etc. These areas have also become ``fiber-less" areas due to their characteristics, where IAB-based systems are likely to be utilized. }
    \label{fig:usecases}
\end{figure*}

\indent Optical fiber is a key-enabler of network deployment in most developed areas, enabling each household to realize high-speed access. However, this requires governments and communities to have a prior line deployment and also requires users to pay directly or indirectly for optical fiber access services. Therefore, unlike developed countries, many regions in developing countries do not have the economic conditions to install optical fiber, which we call ``fiber-less" regions. As shown in Fig.\ref{fig:usecases}, we mainly introduce the following three types of fiber-less regions: (i) rural and remote locations, (ii) low-income neighborhoods in urban and suburban regions, and (iii) post-disaster locations suffering from the destruction of cellular infrastructure. Next, we introduce the characteristics of these three areas in detail. 
\begin{itemize}
    \item \textbf{Rural and remote areas}: Especially in developing countries, rural and remote areas are the most typical ``fiber-less" areas. These areas lack sufficient funds to build base stations and wireless networks. Due to a lower ``potential subscriber density'', a small return on investment (RoI) is one of the main reasons for the lack of investments. Moreover, the cost comes from not only the long fibers needed to reach these areas but also the ``last mile'' cabling under natural conditions such as rivers, mountains, and mud ponds, to name a few. 
    \item \textbf{Low-income neighborhoods in urban and suburban regions}: In urban and suburban regions in developing countries, similar to rural areas, there are some low-income neighborhoods facing economic difficulties. Such areas are typically crowded with unauthorized buildings and little access to resources such as the electrical grid, making it a big challenge for operators to extend fiber connections within these regions.
    \item \textbf{Post-disaster location suffering from the destruction of cellular infrastructure}: The main cause of this type of ``fiber-less" area is natural disasters. Even if these areas once had cellular base stations and fiber optic networks, they cannot be restored quickly after a disaster. Before rebuilding the optical fiber network and base stations, environmental detection and construction planning must be carried out, which makes them ``fiber-less" for a period of time.
\end{itemize}

\section*{IAB networks: virtual fiber}
\indent IAB networks, such as Terragraph and WeLink, mainly operate in the mmWave band to provide consistent, high-bandwidth internet experience \cite{TG,WeLink}. They can be deployed quickly and economically, and they operate best in Line-of-Sight (LoS) to maximize connectivity. The performance of the mmWave channel has been tested in the literature \cite{li2022mobility,du202260}. Terragraph and Cisco both make contributions to their own topology networks. Terragraph is called ``wireless fiber" or ``virtual fiber" because it can support a considerable network capability at a cost lower than that of conventional FWA networks. Digital Alpha Advisors and Cisco invest \$185 million to provide WeLink with a multi-faceted relationship covering a number of fronts. These all indicate that such a topology solution receives not only theoretical support but also market support. \\ 
\indent \textbf{Application:} IAB technique is designed to expand a fiber optic network, it also acts as a high-speed backbone network for multiple local networks. For the next-generation cellular network, the increase in the number of connected nodes and the improvement in service quality are general trends. IAB-based solutions can help mobile operators increase network capacity in fiber-less areas. IAB network is mainly used in residential buildings, apartment buildings, office buildings, school complexes, and any areas where optical fiber connections are not installed, to provide high-speed broadband for fixed users. It can be used in various scenarios such as public safety, smart city, and business services. In the fiber-less areas we mentioned, it can work as a good low-cost solution to achieve high-speed coverage when combined with fixed access connections or Wi-Fi access points. \\
\indent \textbf{Design Method and Preparation:} IAB network is often designed using a mesh topology where there are many different kinds of IAB nodes, such as distribution nodes (DNs) and client nodes (CNs) (here we adopt names in Terragraph, which are named ``anchors" in WeLink). CNs are placed in front of buildings to help customer premise equipment (CPEs) connect to the IAB network, while DNs are placed on mentioned vertical structures. The CPEs (\ie Wi-Fi APs) are directly connected to CNs. When forming their backhaul paths to PoPs, CNs send access requests to their nearby DNs, and such a connection from a CN to a DN is called an access link. In contrast, links between DNs and those between DNs and PoPs are both called backhaul links. CNs and DNs provide backhaul service for CPEs within 15 hops away from PoPs. Since DNs provide both access service for CNs and backhaul service for other DNs, optimizing the spectrum allocation scheme of DNs is a feasible option to maximize network capacity. \\ 
\indent The design of IAB networks requires several steps:
\begin{itemize}
    \item \textbf{Confirm the target location of coverage, \ie understand the market background and the location of customers.} For a certain block or village, the distribution of CPEs is fixed. The installation of CNs is strongly correlated with the distribution of CPEs. When studying network connectivity, the maximum number of supported CNs and the actual data rate of an IAB network can be analyzed by studying the case of network congestion.
    \item \textbf{Determine the location of fiber PoPs.} For each IAB network, there should be at least one PoP connected to the operator's network. The distance between PoPs needs to exceed $250\,{\rm m}$, which is also the maximum distance for DNs to connect to each other. If the distance between PoPs is too small, it will not only bring a lot of costs but also make the IAB network not fully utilized.
    \item \textbf{Site acquisition/Pole identification.} For a typical IAB network, streetlights are the most suitable choice to install DNs, but other vertical structures can also be used, such as signal lights and utility poles. The distribution of such vertical structures in fiber-less regions is random. Among them, the most important thing is to enable each DN to have more than one LoS connection with adjacent DNs. This means that not all DNs' possible positions can be effectively utilized.
\end{itemize}

\indent \textbf{Deployment and Performance testing:} The main goal of IAB networks is to achieve a high-quality network at a low cost. According to the conventional construction process of IAB systems, we propose to follow the below steps to optimize network performance:
\begin{itemize}
    \item \textbf{Pre-installation.} After identifying the location of CPEs, we can get the approximate extent of the IAB network. CNs are generally installed in front of the buildings where CPEs are located. In order to make more CNs form multi-hop connections to PoPs, it is also necessary to pre-install DNs on all possible vertical structures. Next, according to the test of network traffic and capacity, testers can keep the suitable DNs and delete those unsuitable. 
    \item \textbf{Physical connection testing.} To limit the multi-hop delay, DNs that can not form a possible backhaul path to PoPs within 15 hops are possible to be excluded first. This also means that DN density and the distribution of PoPs both affect the proportion of DNs forming possible physical connections to PoPs. In order to allow more users to use IAB network, we need to increase the proportion of DNs with a possible path to PoPs as much as possible in the fiber-less areas we consider. After ensuring that most CNs successfully form physical connections, useless DNs can be found and removed through traffic testing.
    \item \textbf{Network capacity test.} When the locations of PoPs and pre-installed DNs are known, we need to test the maximum number of CNs that the IAB network can support, \ie the network capacity. The peak capacity will not exceed the total capacity of all PoPs. When technicians adopt a unified DNs spectrum allocation framework, it is feasible to optimize the resource allocation of DNs by routing to obtain the maximum actual network capacity. 
    \item \textbf{Minimum proportion of connected DNs:} In order to utilize the capacity of PoPs as much as possible, we need to ensure that for each PoP, there are enough DNs that can help CNs form backhaul paths to it. To save the cost of IAB systems, not all pre-installed DNs can be used eventually, for example, the pre-installed DNs without any traffic may be deleted while the rest can be kept. If there are not enough reserved DNs, it will lead to waste of PoPs capacity. Therefore, it is necessary to test the minimum proportion of DNs connected to PoPs, which makes it possible for the actual network rate to reach the peak. That is, when the distribution and quantity of PoPs are known, at least such a proportion of pre-installed DNs should be reserved so that the network capacity is not always wasted.
\end{itemize}

\section*{Hybrid IAB-based solutions}
\begin{figure*}
    \centering
    \includegraphics[width=1.9\columnwidth]{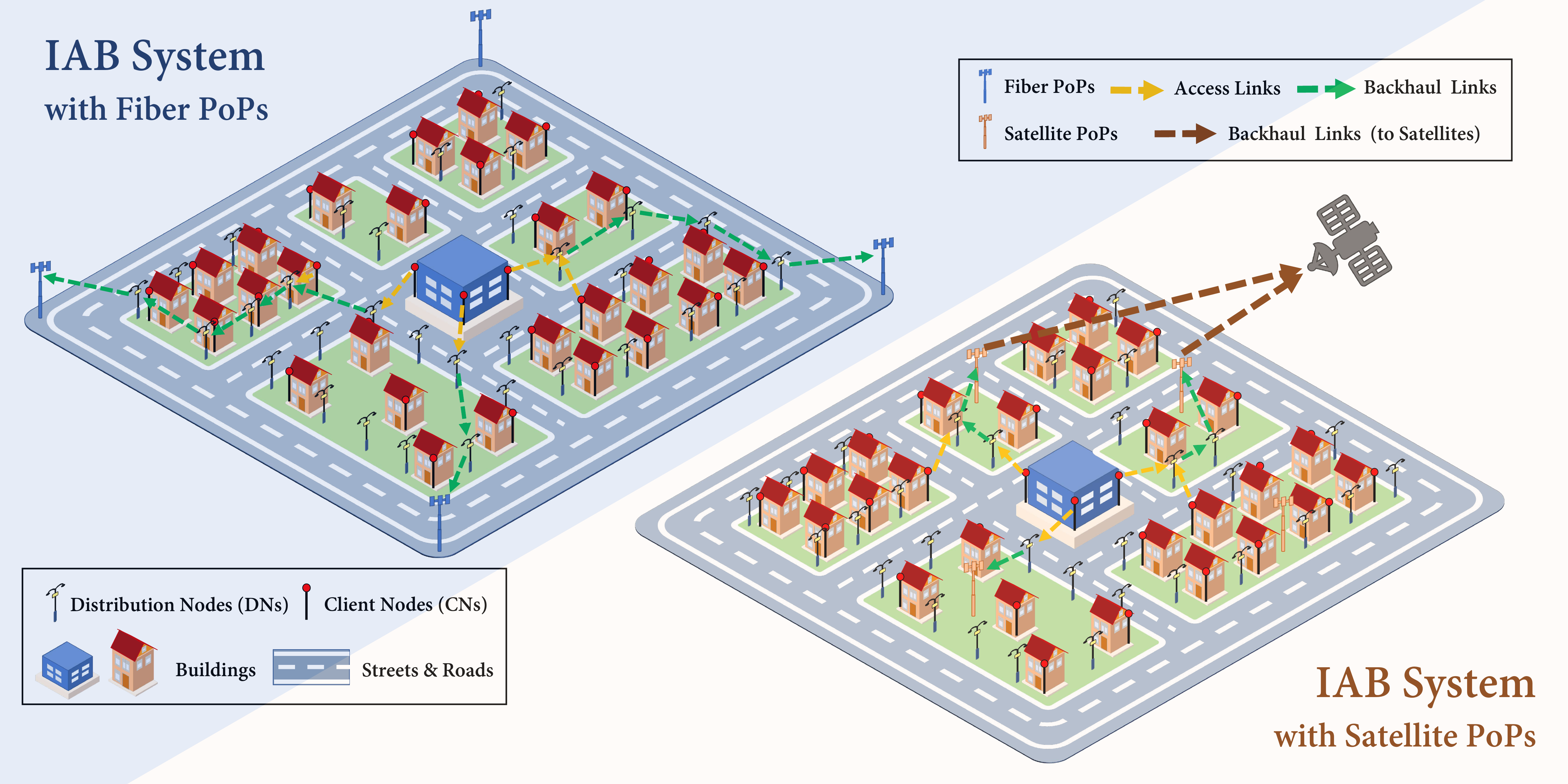}
    \caption{Comparison between conventional IAB network and hybrid IAB-based solutions (taking hybrid IAB + Satellites as an example). Because optical fiber is expensive to build, fiber PoPs in conventional IAB systems would be placed on the boundaries to save the length of optical cables. Different from conventional IAB systems, hybrid IAB systems can break through space limitations using LAPs, HAPs, and satellites. We use a residential block as a demonstration and show the satellites' auxiliary role for the IAB system.}
    \label{fig:TGstrategies}
\end{figure*}
\indent \textbf{FWA and IAB:} FWA and IAB have some similarities, but FWA only provides access services for the last mile area with few optical fibers, while IAB enables the node connection in FWA to be also carried out through mmWave radios. IAB provides a solution for the realization of network-wide fiber-less backhaul.\\
\indent \textbf{Conventional IAB solution}: IAB system (left in Fig.\ref{fig:TGstrategies}) is a novel and low-cost solution that provides fiber-like speed services. It can be a joint radio/free-space optical design for next-generation backhaul systems minimizing the cost. However, it relies on fiber PoPs to connect to its provider's backbone network \cite{douik2016hybrid}. First, low-income areas may not have the capacity to install multiple fiber PoPs. Second, in order to save the expense of long cables between the core network and fiber PoPs, they can be only deployed on the border of residential areas. Therefore, it is difficult for them to serve users in different locations in fiber-less areas.\\
\indent \textbf{Hybrid IAB + NTN solution:} To enhance global connectivity, many NTN platforms, such as balloons and satellites, are proven viable to assist mmWave networks, especially backhaul networks \cite{yaacoub2019key,sanchez2020millimeter,bertizzolo2019mmbac}. Based on the IAB solution, the use of NTN platforms makes up for the spatial limitation of PoPs caused by insufficient funds and further saves the possible cost, which is called hybrid IAB + NTN solutions (right in Fig.\ref{fig:TGstrategies}). However, such strategies would require access to the NTN platforms.\\ 
\indent {In this paper, we propose to use IAB and hybrid IAB-based solutions to enhance the connectivity of fiber-less areas. In the following section, we introduce concrete use cases of IAB solutions to support our concept.} 

\section*{Use cases and proof of concept} \label{usecases}
\indent Based on the mmWave spectrum, Line-of-Sight (LoS) is a great help to establish stable links between IAB nodes, especially in rural areas. Several vertical structures, such as streetlights and utility poles, are suitable for installing DNs. LoS is also often achieved between these vertical structures. There are different vertical structures in different types of fiber-less areas, that have brought opportunities for the construction of IAB networks.\\
\noindent {\em Use Case 1: Rural and remote locations.}\\
\indent In rural and remote areas, wireless networks can help enhance smart agriculture and environmental monitoring. Rural areas generally range in size from $1\,{\rm km^2}$ to $10\,{\rm km^2}$. Especially in flat farms and pastures, because there is no obstruction of high-rise buildings, most signals can be transmitted through LoS. But if the nodes are set in forests or mountains, the probability of LoS will decrease. At this time, better transmission performance can only be obtained by manually adjusting the position of DNs. Different from those in urban areas, DNs in rural areas may not only rely on streetlights but may also be set on more different vertical structures, like wind turbines (WTs), meteorological stations, and solar insecticidal lamps Internet of Things (SIL-IoTs). Extensive mmWave spectrum experiments in agricultural fields have been conducted to study the path loss in these areas \cite{vuran2022millimeter}.\\
\noindent {\em Use Case 2: Low-income neighborhoods.}\\
\indent In low-income neighborhoods in urban or suburban regions, backhaul networks can help with remote learning, smart transportation, and many other services needed by the people living in such neighborhoods. Such a neighborhood is often part of an area larger than $10\,{\rm km^2}$. Since the buildings are not very high, it is possible to achieve LoS between most DNs. Due to the trend of urbanization in these areas, DNs can rely on various vertical structures such as utility poles and streetlights, and these structures can realize LoS between most adjacent nodes. The performance of mmWave channels in streets and residential areas was also tested, such as in the work on \cite{aslam2020analysis}.\\
\noindent {\em Use Case 3: Post-disaster locations.}\\
\indent The size of post-disaster areas is difficult to measure, but most of them have lost high buildings and base stations. IAB networks can help them detect the environment and communicate to speed up reconstruction. Due to the need to restore lighting and power to these areas in time, a number of temporary but stable vertical structures are installed to assist in the safe and rapid reconstruction work on the ground. These structures like streetlights and utility poles can also achieve LoS between most adjacent nodes easily. Due to the ease of installation, DNs can be assembled with the help of these structures and the communication system can be quickly restored.\\
\indent {It is worth noting that, both the conventional IAB solution and hybrid IAB + NTN solution can not only enhance the connectivity of the mentioned fiber-less areas but also work as the complement of existing communication networks. Therefore, IAB-based solutions can not only connect the unconnected areas, but also super-connect the connected areas.}

\section*{Simulations} \label{sec:sim}
\begin{figure}
    \centering
    \includegraphics[width=1\columnwidth]{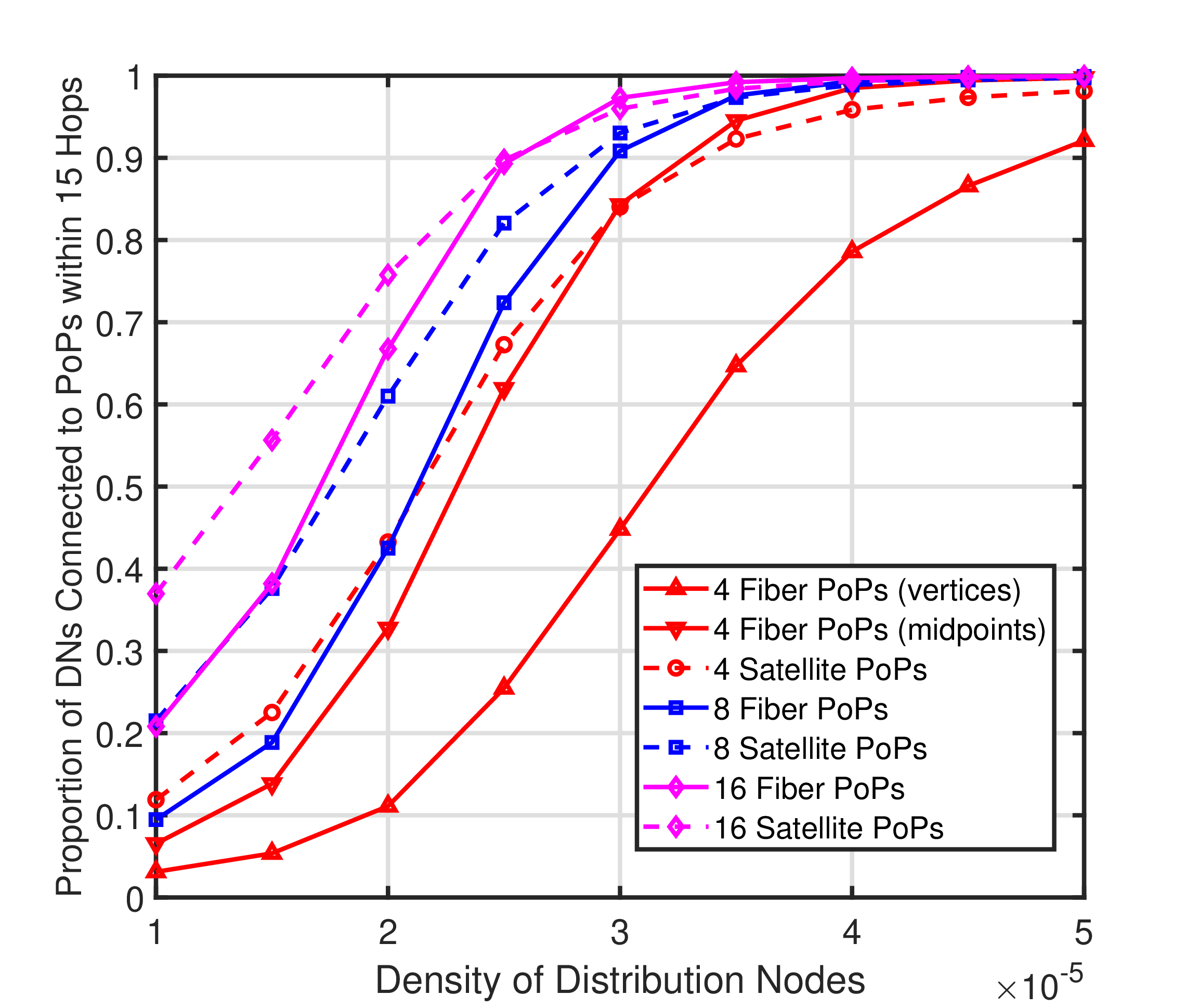}
    \caption{The relationship between the density of pre-installed DNs and the proportion of DNs connected to PoPs within 15 hops.}
    \label{fig:LGP_lamTG}
\end{figure}

\begin{figure}
    \centering
    \includegraphics[width=1\columnwidth]{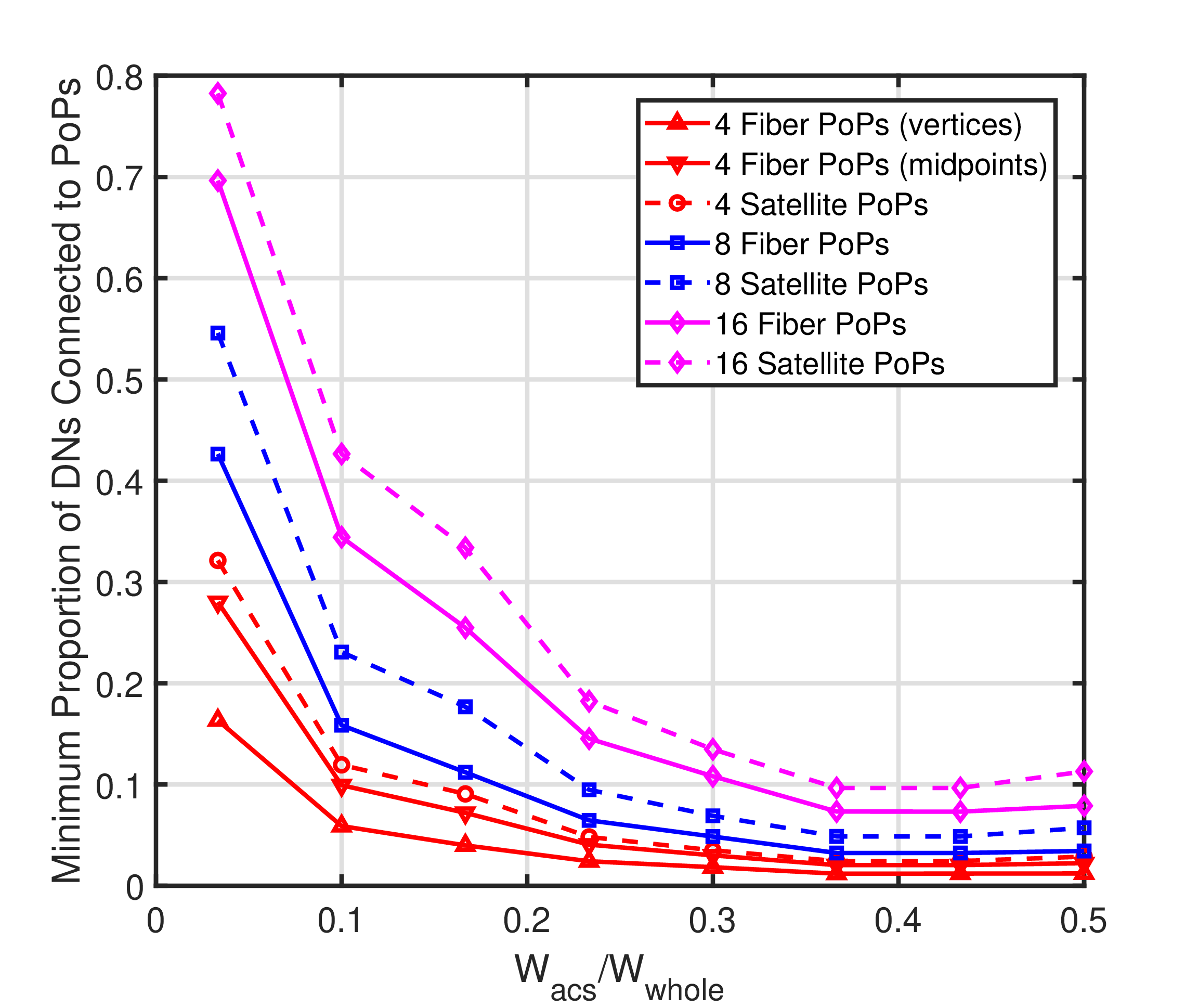}
    \caption{The relationship between the proportion of bandwidth that access service occupies in each DN's spectrum and the minimum proportion of DNs connected to PoPs. We set the density of pre-installed DNs is $5\times 10^{-5} \,{\rm DNs/m^2}$ in a $\rm 4000\,{m}\times 4000\,{m}$ area.}
    \label{fig:MinPro_TG}
\end{figure}

\begin{figure}
    \centering
    \includegraphics[width=1\columnwidth]{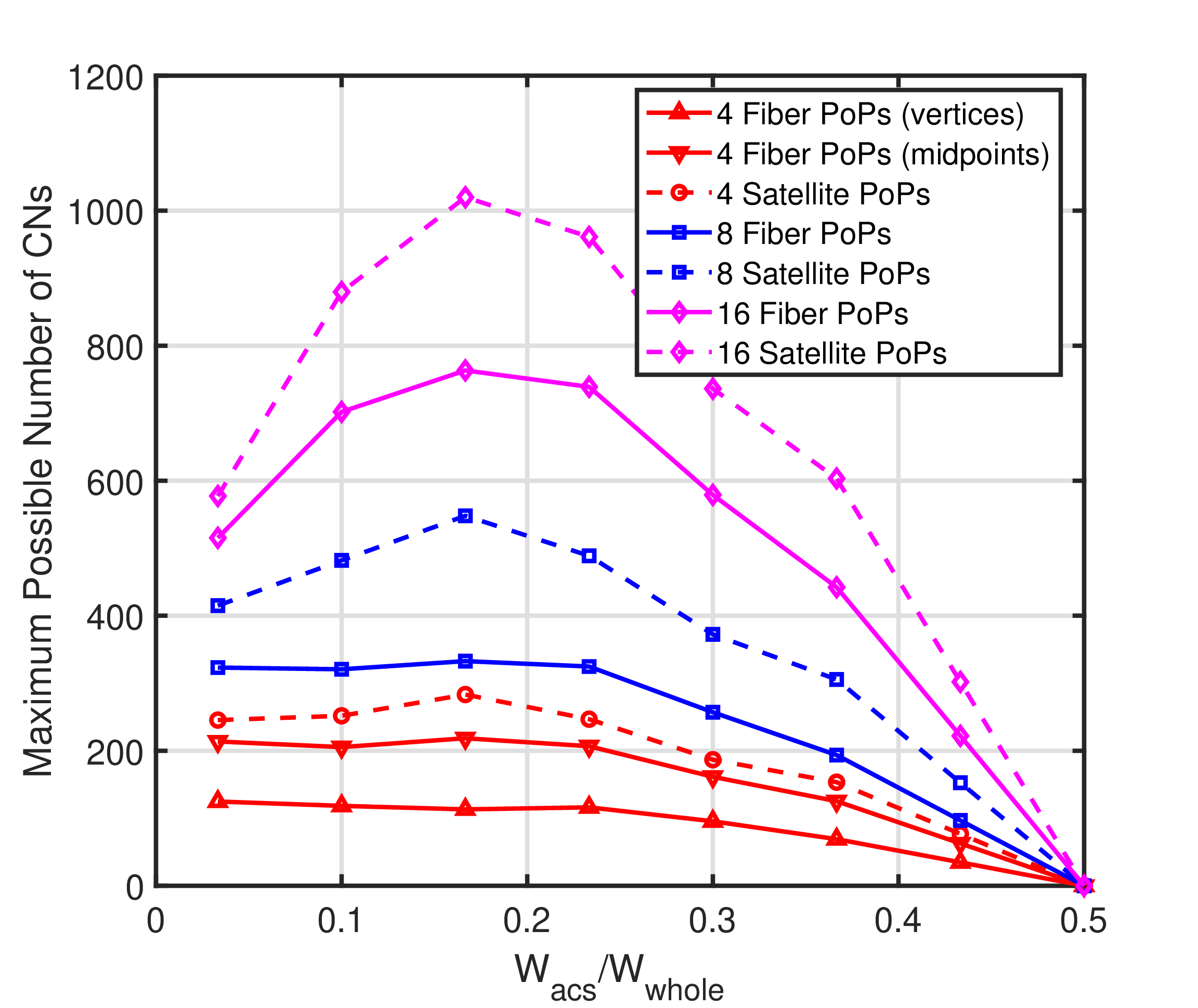}
    \caption{The relationship between the proportion of bandwidth that access service occupies in each DN's spectrum and the maximum possible number of CNs. The density of pre-installed DNs is $5\times 10^{-5}\,{\rm DNs/m^2}$ in a $\rm 4000\,{m}\times 4000\,{m}$ area. }
    \label{fig:MaxNum_AP}
\end{figure}

\indent {In order to investigate the feasibility of the conventional IAB solution and the advantages of the proposed hybrid IAB + NTN solution, we build a $4\,{\rm km}\times 4\,{\rm km}$ square residential area for {MATLAB} simulation. We mainly consider LEO satellites in the altitude range of 160-2000 ${\rm km}$ as relay nodes between PoPs and the core network, whose transmit power is commonly assumed as $15\,{\rm dBW}$. The channels between satellite PoPs and LEO satellites are typically assumed to follow shadowed Rician fading models. In the following, we use the term satellite PoP to refer to the PoPs that are connected to LEO satellites. Note that HAP/LAP PoPs would have the same influence since we are only considering the case where the NTN-connected PoP has reliable connections to the core network.}\\
\indent We take fiber PoPs and satellite PoPs as examples to discuss network performance under different PoPs distribution schemes. Especially when there are 4 fiber PoPs, we study two cases where the fiber PoPs are located at the vertices of the block and at the midpoints of the boundaries. {Inspired by realistic scenarios, some important simulation parameters, are assumed to carry the values listed in Table \ref{tab:TableOfNotations}}. We use $W_{\rm acs}/W_{\rm whole}$ to represent the proportion of bandwidth that access service occupies in each DN's spectrum. \\
\begin{table}[t]\caption{Parameters in Simulation}
\centering
    \begin{tabular}{ {l} | {l} }
    \hline
        \hline
    \textbf{Parameter} & \textbf{Value} \\ \hline
    Simulation range & $4\,{\rm km}\times 4\,{\rm km}$\\ \hline
    The capacity of each PoP & $5\,{\rm Gbps}$\\ \hline
    The capacity of each DN & $1.5\,{\rm Gbps}$\\ \hline
    The capacity each CN occupies & $50\,{\rm Mbps}$\\ \hline
    Maximum communication range between adjacent DNs & $250\,{\rm m}$\\ \hline
     \hline
    \end{tabular}
\label{tab:TableOfNotations}
\end{table}
\indent First, we study the relationship between the density of pre-installed DNs (DN density) and proportion of pre-installed DNs forming multi-hop links (\ie backhaul paths) to PoPs within 15 hops. As shown in Fig.\ref{fig:LGP_lamTG}, when the DN density is less then $2\times 10^{-5}\,{\rm DNs/m^2}$, satellite PoPs uniformly distributed inside the block can make more DNs connect to PoPs than when we adopt fiber PoPs. When we increase the DN density, a higher proportion of pre-installed DNs can form backhaul paths to PoPs. When the DN density is larger than $3\times 10^{-5}\,{\rm DNs/m^2}$, most pre-installed DNs can form backhaul paths (multi-hop links) to PoPs. Except for the case where 4 fiber PoPs are located on the vertices, the proportions of connected DNs in other strategies are more than $80$ percent. In addition, the fiber PoPs at midpoints of the sides have better performance than those at vertices, which is in agreement with the concept in \cite{TG}.\\ 
\indent As observed from Fig.\ref{fig:LGP_lamTG}, sufficient DN density ensures the capability of majority of the deployed DNs to reach a PoP within less than 15 hops. However, due to limitation of resources available at the PoPs, particularly available bandwidth, only a proportion of the deployed DNs are needed to achieve full use of the available PoP resources. If the deployed DN density is less than this proportion, the PoP resources are considered wasted. On the other hand, if more DNs are deployed than this proportion, not all DNs will be used due to resources limitation at the PoPs. Hence, to achieve a cost-efficient deployment, the density of deployed DNs should be as close as possible to this minimum density (which is minimum proportion times the given density) to ensure full use of deployed DNs and full use of PoP resources. In Fig.\ref{fig:MinPro_TG}, for DN density $=5\times10^{-5}\,{\rm DNs/m^2}$, we show how this proportion varies as we increase the bandwidth allocated for access services. When we increase the spectrum occupied by the access service, such minimum density decreases. When the amount and capacity of PoPs are fixed, satellite PoPs can support more DNs successfully than fiber PoPs.\\
\indent When the number and capacity of PoPs are known, the actual network capacity mainly depends on the ability to form multi-hop backhaul links. Considering a unified DN spectrum allocation method, we obtain the optimal fixed spectrum allocation scheme. As shown in Fig.\ref{fig:MaxNum_AP}, when the proportion of access spectrum to the whole spectrum of each DN is between $0.1$ and $0.25$, the maximum network capacity can be obtained. {For example, an IAB network with 16 satellite PoPs can support at least 1000 CNs when $W_{acs}/W_{whole}=1/6\approx 1.67$, where the throughput of the whole IAB network is 50 Gbps}. When the access spectrum is half of the total spectrum, only the CNs near one-hop DNs can be supported stably.\\

\section*{Challenges} \label{sec:challenge}
\indent \textbf{Challenge 1.} \textbf{Lack of incentives and regulation.} As a possible solution for fiber-less areas, it still needs to be approved by the authorities. At the same time, during the period of use, IAB tech still need professional technicians to maintain the management of the backhaul networks, which also imposes requirements on the service of the operator.\\
\indent \textbf{Challenge 2.} \textbf{Pre-deployment strategies of different scenarios.} In this paper, we introduce several IAB-based solutions, including conventional IAB systems and hybrid IAB-based backhaul systems. Although the hybrid IAB-based solutions can break through the limitation of space to obtain better actual network capacity, the feasibility of these solutions is ultimately determined by the authorities and regulators, which consider many problems including the willingness to add or move fiber PoPs, the support from satellites, LAPs or HAPs, etc. Different solutions bring different opportunities and challenges, which make it difficult for their decision-making.\\
\indent \textbf{Challenge 3.} \textbf{Network design optimization.} The network performance analysis in this paper includes i) the characteristics of DNs with a unified spectrum allocation and ii) the optimal spectrum allocation of DNs, especially those directly connected to PoPs, \ie one-hop DNs. But for the overall network, it is also feasible to further adopt individualized allocation or adaptive allocation, but they are more difficult and expensive to realize. At the same time, in order to ensure the stable operation of IAB networks, the operators often require that each DN must be connected to more than two neighbor nodes to deal with sudden DN damage. {In addition, adaptive routing algorithms and the assistance of NTN networks can also help promote network resilience of our proposed IAB-based solutions.} \\
\indent \textbf{Challenge 4.} \textbf{LoS.} In our study, LoS is an important condition for stable connections between DNs, especially in rural areas. Most fiber-less areas we consider satisfy the LoS condition, and they are possibly better than developed cities. But due to the influence of terrain, road signs, and buildings, two adjacent DNs may still be unable to connect, which brings difficulties to network construction. \\
\indent \textbf{Challenge 5.} \textbf{Energy efficiency.} Even though the cost of IAB equipment is cheaper than optical fiber, the daily access and backhaul services still bring a lot of energy demand. {Vertical structures like streetlights can help DNs connect to the electricity grid easily. Solar power strategy can be a feasible method to support DNs without increasing the burden on the electricity grid. In realistic applications, routing is also a main factor in increasing energy requirements, while many adaptive routing algorithms have been proposed to reduce the energy consumed.} \\
\indent \textbf{Challenge 6.} \textbf{Decrease in node capacity.} Due to the real-time routing requirements in multi-hop communication, the capacity of each DN may be fully utilized such that it can not support other DNs by forwarding their packets, which also leads to a decrease in the number of CPEs and CNs that can actually access the backhaul system.\\
\indent \textbf{Challenge 7.} \textbf{User fairness.} User fairness in IAB systems means that users in different locations in the same residential area hope to be provided with the same and good quality of service (QoS). How to satisfy this requirement is an important problem, while routing and spectrum management are both possible solutions.\\
\indent \textbf{Challenge 8.} \textbf{Vertical integration.} {In the context of integrated ground-air-space networks, vertical integration of resources between terrestrial network (TN) operators and NTN operators is particularly important. A flexible vertical management mechanism can not only enable the network throughput but also save the cost for the proposed hybrid IAB + NTN solution. However, it still faces many challenges, such as market competition and resource allocation.}
\section*{Conclusion} \label{sec:conclusion}
\indent In this article, we have discussed the characteristics and development needs of ``fiber-less" regions. We have introduced solutions for high-speed low-cost services based on IAB systems. Conventional IAB systems rely on fiber PoPs, whose capacity is difficult to be fully utilized in terms of topology, especially in low-income areas, while the hybrid IAB systems rely on new NTN platforms such as HAPs, LAPs and satellites. For the ``fiber-less" areas, we have verified that sufficient pre-installed DN density establishes the paths to PoPs for most DNs. The topology flexibility of NTN platforms makes the backhaul paths be constructed more efficiently. That means the hybrid solution of IAB and NTN reduces the cost of link testing. We mainly consider the case where all DNs adopt unified spectrum pre-allocation, and verify the minimum proportion of reserved DNs under different spectrum allocation schemes, which is also the minimum proportion of pre-installed DNs that PoPs can serve in case of network congestion, to optimize the actual network capacity. The lower the proportion of the access spectrum in the overall spectrum, the more DNs can work simultaneously. But only when it is between 0.1 and 0.25, the most CNs can be connected to the core network. Finally, we also discuss the challenges faced by such IAB solutions.

\ifCLASSOPTIONcaptionsoff
  \newpage
\fi

\bibliographystyle{IEEEtran}
\bibliography{ref}


\end{document}